\documentclass[aps,physrev,preprint,superscriptaddress]{revtex4-2}

\usepackage{physics, comment}
\usepackage{subcaption}
\usepackage{graphicx}
\usepackage{appendix}
\graphicspath{{figures/}}

\usepackage[colorlinks=true,linkcolor=red,urlcolor=blue,citecolor=blue]{hyperref}

\begin{document}
\title{Time-Averaged Template for Stochastic Gravitational-Wave Background Detection in Space-Based Interferometers}

\author{Jing-yi Wu}
\affiliation{School of Astronomy and Space Science, University of 
Chinese Academy of Sciences (UCAS), Beijing 100049, China}

\author{Yong Tang}
\affiliation{School of Astronomy and Space Science, University of 
Chinese Academy of Sciences (UCAS), Beijing 100049, China}
\affiliation{School of Fundamental Physics and Mathematical Sciences,
 Hangzhou Institute for Advanced Study, UCAS, Hangzhou 310024, China}
\affiliation{International Center for Theoretical Physics Asia-Pacific, Beijing 100190, China}

\begin{abstract}
Stochastic gravitational-wave background (SGWB) poses significant challenges for data analysis and parameter inference in future space-based gravitational-wave missions, such as LISA and Taiji, as it appears as an additional stochastic component along with instrumental noise. Previous studies have developed various approaches to distinguish the SGWB from instrumental noise, often under simplified assumptions such as static or equal-arm configurations. However, in realistic scenarios, time-varying arm-lengths introduce additional complexities that require careful modeling. 
In this work, we investigate the impact of template construction on SGWB parameter estimation under realistic orbital configurations. Using the simulated SGWB signals and dominant instrumental noise sources, we compare three template strategies: time-averaged template constructed from segmented data, equal-arm template, and a template treating the arm-lengths as a free parameter. Our results show that the time-averaged template yield improves parameter estimation accuracy under time-varying arm-lengths, whereas introducing the effective arm-length as a free parameter increases estimation uncertainty. These findings highlight the importance of realistic template construction for high-precision SGWB analysis in future space-based missions.
\end{abstract}

\maketitle
\newpage

\section{Introduction}
Since the first detection of gravitational waves (GWs) in 2015 \cite{PhysRevLett.116.061102}, the accumulated merger events have significantly advanced our understandings in astrophysics and gravitational physics. To extend gravitational-wave observations to lower frequency bands, which offer deep and extra insights into the universe, space-based laser interferometers, such as LISA \cite{amaroseoane2017laserinterferometerspaceantenna}, Taiji \cite{10.1093/nsr/nwx116}, and TianQin \cite{Luo_2016}, are scheduled to operate in the mid-2030s.

One of the primary scientific goals of these space-borne GW detectors is the detection of the SGWB. The origins of the SGWB can be broadly categorized into two types: astrophysical sources and the cosmological background. The astrophysical SGWB originates from the superposition of numerous weak GW signals, primarily emitted by binary compact objects, such as stellar-origin black hole (BH) mergers and neutron star binaries \cite{Regimbau_2006,Regimbau_2007,phinney2001practicaltheoremgravitationalwave,Regimbau_2011,Marassi_2011,babak2023stochasticgravitationalwavebackground} or extreme mass ratio inspirals \cite{Bonetti_2020,pozzoli2023computationstochasticbackgroundextreme}. The cosmological SGWB can arise from various mechanisms, including vacuum fluctuations \cite{Starobinsky:1979ty,Fabbri:1983us,Allen:1987bk,Seljak_1997,Kamionkowski_1997,Ananda_2007}, preheating during the post-inflation period \cite{Easther_2006,Easther_2007,Garc_a_Bellido_2007,Dufaux_2007,Garc_a_Bellido_2008,Bethke_2013,Bethke_2014}, first-order phase transitions \cite{Kamionkowski_1994,Apreda_2002,Grojean_2007,Caprini_2008,Huber_2008,Caprini_2009,Caprini:2009,Leitao_2012,Hindmarsh_2014,Giblin_2014,Kakizaki_2015,Leitao_2016,Jinno_2017,Chao_2017,Weir_2018,Chala_2018,Hindmarsh_2019,liang2025bayesiananalysiscomplexsinglet,guan2025measuringgravitationalwavespectrum}, and cosmic defects \cite{PhysRevD.31.3052,PhysRevD.42.354,PhysRevD.45.3447,Battye_1994,Caldwell_1996,Figueroa_2013,Blanco_Pillado_2017,Liu_2021}; for comprehensive discussions, see reviews \cite{Caprini_2018,Auclair_2023}. The detection of the cosmological SGWB would provide critical knowledge about the early universe, while searches for the astrophysical SGWB enhance our understanding of the population and evolution of binary compact objects.

Due to its stochastic nature, the SGWB manifests as an additional noise component, mixing with instrumental noise and posing significant challenges for GW data analysis. This problem has driven extensive research into methods for distinguishing the SGWB from instrumental noise. Ref.~\cite{Adams_2010} intends to disentangle the SGWB using combinations of time-delay interferometry (TDI) channels, focusing on fixed spectral shapes for both the signal and noise.
However, data from the LISA Pathfinder mission \cite{PhysRevLett.120.061101,PhysRevLett.116.231101} indicates that observed noise levels at low frequencies can deviate from simple expectations, emphasizing the necessity of modeling flexible noise sources. To address unknown or non-parametric spectra, several studies have investigated flexible models for both instrumental noise and the SGWB \cite{Caprini_2019,Flauger_2021,Baghi_2023,muratore2023impactnoiseknowledgeuncertainty,Pozzoli_2024,Caprini_2024,blancopillado2025gravitationalwavescosmicstrings,braglia2024gravitationalwavesinflationlisa,gammal2025reconstructingprimordialcurvatureperturbations,santini2025flexiblegpuacceleratedapproachjoint,aimen2025bayesianpowerspectraldensity}. Other works have characterized the SGWB and instrumental noise by allowing for unequal noise levels across the optical links \cite{Adams_2010,Adams_2014,Wang_2022,Hartwig_2023,Wang_2024,Kume_2025}.
Most of these studies, however, relied on simulated data assuming static orbital configurations or employed the equal-arm approximation during parameter estimation. In reality, arm-length variations caused by the detector’s orbital motion introduce additional uncertainties in signal extraction. To address these variations and other realistic challenges—such as data gaps and complex noise artifacts—various data challenge initiatives have been launched, including the LISA Data Challenge (LDC) \cite{baghi2022lisadatachallenges}, the Taiji Data Challenge (TDC) \cite{Ren_2023,du2025realisticdetectionpipelinestaiji} and the GWSpace project~\cite{Li_2025}. The generation of these realistic simulated datasets has spurred the development of advanced parameter estimation methods.
To mitigate the impact of arm-length variations, the Short-Time Fourier Transform (STFT) has been applied to analyze various GW sources, such as galactic binaries, massive black hole binaries, and stellar-mass black hole binaries, as well as for noise characterization \cite{Digman_2022,Digman_2023,du2025enhancingtaijisparameterestimation}. Furthermore, this approach has been utilized to simultaneously extract both isotropic and anisotropic SGWB signals from instrumental noise \cite{criswell2025flexiblespectralseparationmultiple}. While previous studies have utilized time-frequency analysis methods such as STFT, this approach is excessively demanding on computational resources. This high demand stems from the parameter estimation process, which requires the computation and inversion of the frequency-domain correlation matrix across multiple channels. It is particularly resource-intensive for non-orthogonal channels, such as the Michelson combination. 

Here, we investigate the impact of template construction on parameter estimation under realistic orbital conditions. To improve efficiency, we instead perform the analysis directly in the frequency domain. We generate data that includes the SGWB and two primary instrumental noises: test mass acceleration noise (acc) and optical metrology system noise (oms).
In this initial study, we construct the model by assuming fixed power spectral density (PSD) shapes for all three components, and that the two noise sources are identical across the six optical benches.
To assess the influence of template on high-precision signal characterization, we compare three distinct template strategies. 
The first is the time-averaged template, obtained by segmenting the entire time series into shorter durations, calculating the frequency-domain template for each segment, and subsequently averaging the results.
The second is the idealized equal-arm template. 
The third template treats the arm-length $L$ from the equal-arm model as an additional free parameter, allowing us to test the effect of increased model complexity on the final signal extraction.
To mitigate statistical uncertainties and facilitate comparison, we generate and analyze datasets with durations of one year, a half-year, and three months. Our results indicate that for orbits featuring time-varying arm-lengths, employing the time-averaged template significantly reduces parameter estimation errors compared to the idealized equal-arm approximation template. Conversely, introducing the arm-length $L$ as an additional free parameter leads to even larger estimation errors. Therefore, for high-precision parameter estimation under realistic orbital conditions, it is advised to segment long-duration data to compute frequency-domain template, thereby improving the overall accuracy of the estimation.

The remainder of this paper is organized as follows. Section~\ref{sec:datasimulation} describes the data simulation methods, including the characterization of the injected SGWB and instrumental noise, as well as the detailed parameters of the simulated datasets.
In Section~\ref{sec:data_analysis}, we present the formulation of the posterior distribution and the derivation of the three analysis templates. Furthermore, we show a comparison of the obtained results.
Section~\ref{sec:results} is dedicated to calculating the Bayes factor to investigate the preference for the different models. The conclusions are summarized in Section~\ref{sec:conclusion}.

\section{Theoretical Formalism and Data Simulation} \label{sec:datasimulation}
In this section we establish the theoretical framework and conventions. We characterize the SGWB and the instrumental noise with their analytical PSD, and elaborate how we implement in the data simulation.

\subsection{Characterization of SGWB}
The total gravitational wave field $h(t, \vb{x})$ in the transverse-traceless (TT) gauge, expressed as a superposition of many plane waves with different frequencies $f$ and propagation directions $\vb{\hat{k}}$, is given by
\begin{equation}
    h(t,\vb{x})=\int_{-\infty}^{+\infty}\int_{\vb{\hat{k}^2}} \left[ \tilde{h}_{+}(f;\vb{\hat{k}})\epsilon_{+}+\tilde{h}_{\times}(f;\vb{\hat{k}})\epsilon_{\times} \right]e^{i2\pi f \left(t-\vb{\hat{k}\cdot\vb{x}}\right)} \dd[2]{\vb{\hat{k}}}\dd{f}.
    \label{eq:planewave}
\end{equation}
Here $\epsilon_{+}$ and $\epsilon_{\times}$ are the ``plus'' and ``cross'' polarizations.

Since the SGWB inherently behaves as another source of noise, analyzing its properties directly in the time domain is often inconvenient, especially when large stochastic instrumental noise components are present. A more effective approach is to convert the time-series data into PSD in the frequency domain.
For simplicity, we first consider the case where the SGWB is stationary, Gaussian, isotropic, and unpolarized. 
Then the one-sided PSD of the gravitational wave strain, $S_{\text{gw}}(f)$, is defined by the two-point correlation function of the Fourier-transformed strain components \cite{Allen_1999}
\begin{equation}
    \expval{ \tilde{h}_A(f;\vb{\hat{k}})\tilde{h}^*_{A'}(f';\vb{\hat{k}'}) }_\text{ens} = \frac{1}{16\pi} S_{\text{gw}}(f) \delta(f-f')\delta_{AA'}\delta_{\vb{\hat{k}}\vb{\hat{k}'}} 
    = \frac{T_{\text{obs}}}{16\pi} S_{\text{gw}}(f)\delta_{AA'}\delta_{\vb{\hat{k}}\vb{\hat{k}'}},
    \label{eq:psd}
\end{equation}
where $A={+,\times}$ represents the polarization and $T_{\text{obs}}$ is the observation time. The angle bracket $\expval{...}_{\text{ens}}$ denotes the ensemble mean. 
The constant $16\pi$ arises from three components: $4\pi$ accounts for the average over all sky directions, 2 accounts for the average over the two polarizations, and the final factor of 2 accounts for the conversion from the one-sided PSD to the two-sided spectrum. 
Therefore, considering the summed strain from all directions in the sky and all polarizations, the resulting one-sided PSD is
\begin{equation}
    \expval{\tilde{h}(f)\tilde{h}^*(f)}_{\text{ens}}=\frac{T_{\text{obs}}}{2} S_{\text{gw}}(f).
    \label{eq:psd_sum}
\end{equation}

The GW strain PSD $S_{\text{gw}}(f)$ is related to the dimensionless GW energy density spectrum $\Omega_{\text{gw}}(f)$, as \cite{Allen_1999}
\begin{equation}
    S_{\text{gw}}(f) = \frac{3 q_0^2}{2 \pi^2 f^3}\left[q^2\Omega_{\text{gw}}(f)\right],
\end{equation}
where $q$ is the dimensionless Hubble parameter, defined such that $H_0 = q \cdot q_0$, and $q_0=100~\rm{km}\cdot\rm{s}^{-1}\cdot\rm{Mpc}^{-1}$ is the reference constant.

The spectrum we consider for the SGWB has a single power-law shape, which is commonly employed in modeling GWs originating from binary stellar-mass BH systems or neutron stars and some of the inflationary models~\cite{Caprini_2019,Flauger_2021,Dimitriou_2024}. This spectrum is parameterized by two physical quantities: the energy density amplitude $\Omega_0$ and the spectral index $\gamma$. The dimensionless energy density spectrum is thus described as
\begin{equation}
    q^2\Omega_{\rm{gw}}(f) = \Omega_0 \left( \frac{f}{f_0} \right)^\gamma.
    \label{eq:singlepw}
\end{equation}
The pivot frequency $f_0$ is fixed to $1~\text{mHz}$ in the our simulation. 
As $f_0$ serves merely as a scaling parameter, its choice does not impact the signal reconstruction.

\begin{figure}[t]
    \centering
    \includegraphics[width=10cm, height=8cm]{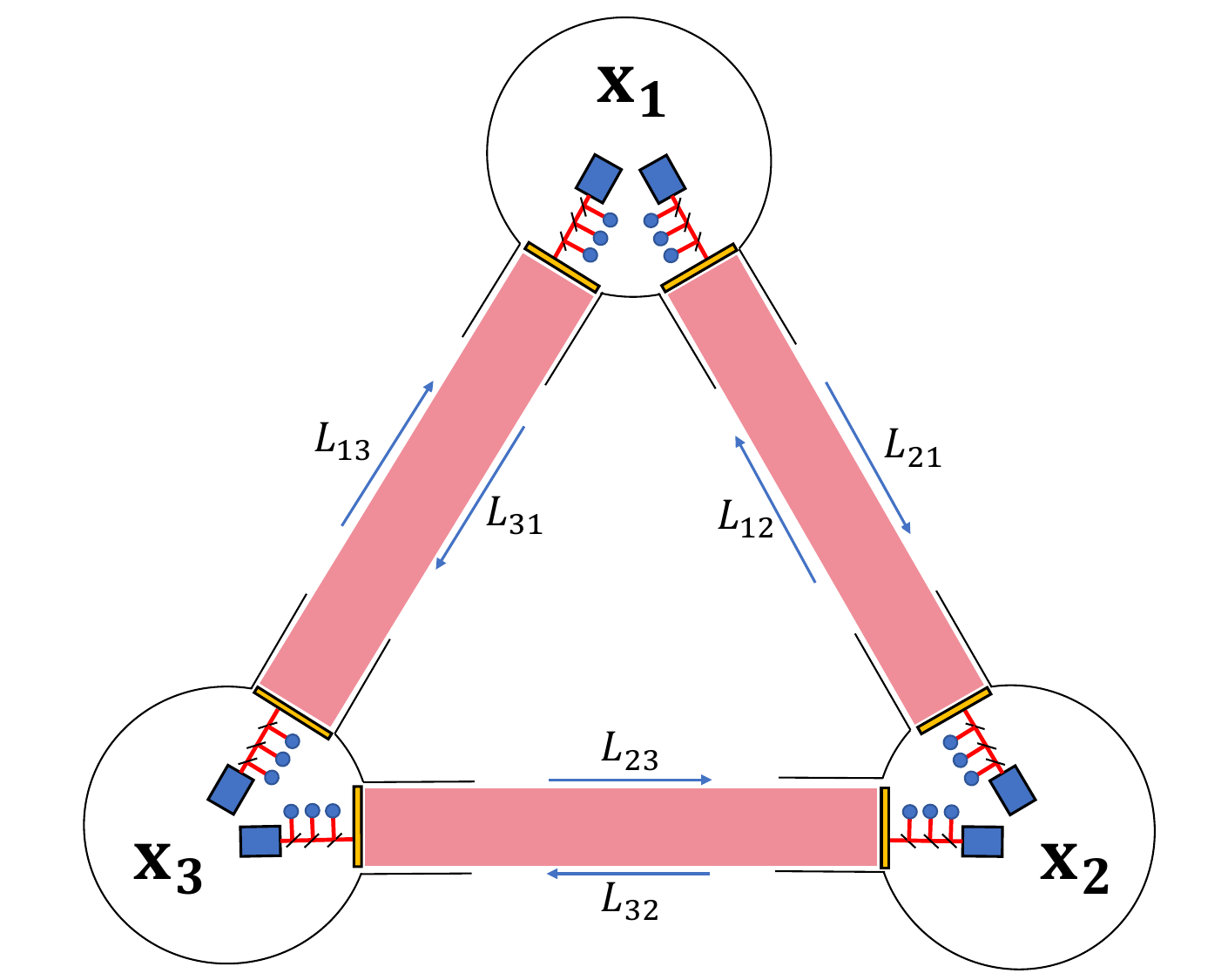}
    \caption{The schematic diagram of the links for Taiji, which orbits around the Sun. The triangular constellation consists of three spacecraft labeled $\vb{x}_1$, $\vb{x}_2$, $\vb{x}_3$. Each spacecraft is equipped with two optical benches designed to monitor distance changes, thereby defining six single-link interferometric arms. The subscript of the light travel time $L_{\text{rs}}$ denotes the specific direction of the link, running from the sender $\vb{x}_\text{s}$ to the receiver $\vb{x}_\text{r}$.}
    \label{fig:diagram}
\end{figure}

The raw data recorded by laser interferometers are expressed as the fractional frequency shift induced by the Doppler effect of GW perturbations on the metric. 
The single-link signal is parameterized as \cite{Baghi_2023}
\begin{equation}
    y_{\text{rs}}(t;\vb{\hat{k}})=\frac{1}{2\left[1-\vb{\hat{k}} \cdot \vb{\hat{n}_{\text{rs}}}(t)\right]}\left[H_{\text{rs}}\left(t-L_{\text{rs}}(t)-\vb{\hat{k}} \cdot \vb{x_s}(t)\right)-H_{\text{rs}}\left(t-\vb{\hat{k}} \cdot \vb{x_r}(t)\right) \right], \label{eq:gwfreshift}
\end{equation}
where the subscript $\text{rs} \in \left\{12, 23, 31, 13, 32, 21\right\}$ denotes the specific laser link. $\vb{\hat{n}_{\text{rs}}}$ is the unit vector pointing from the sender spacecraft $\vb{x}_\text{s}$ to the receiver spacecraft $\vb{x}_\text{r}$, and $L_{\text{rs}}$ is the corresponding light travel time along the arm. Furthermore, $H_{\text{rs}}$ is the projected strain along the arm rs, which is defined by the inner product of the arm unit vector and the strain tensor, $H_{\text{rs}}=\hat{n}^i_{\text{rs}}\hat{n}^j_{\text{rs}}h_{ij}$. For a more intuitive illustration of these laser link symbols and the overall geometry, a schematic diagram of the Taiji constellation is shown in Fig.~\ref{fig:diagram}. 
 
Consequently, the total single-link signal $y_{\text{rs}}(t)$, which represents the full response to the stochastic background, is the superposition of the contributions from all directions $\vb{\hat{k}}$
\begin{equation}
    y_{\text{rs}}(t)=\sum_{\vb{\hat{k}}}y_{\text{rs}}(t;\vb{\hat{k}}).
\end{equation}
For the simulations performed in this study, the full sky is divided by 108 uniformly spaced angular regions, each representing a distinct propagation direction $\vb{\hat{k}}$.

\subsection{Characterization of the instrumental noise}
The total instrumental noise in the laser links consists of various components. In this work, we focus on the two dominant contributors for space-based interferometers: the test mass acceleration noise $n^{\text{acc}}_{\text{rs}}(t)$ and the noise of optical metrology system $n^{\text{oms}}_{\text{rs}}(t)$. The one-sided PSD for them are modeled as \cite{amaroseoane2017laserinterferometerspaceantenna}
\begin{equation}
    S_{\text{acc}}(f)=\left( \frac{s_{\text{acc}}}{2\pi fc} \right)^2 \left[1+\left(\frac{0.4~\rm{mHz}}{f} \right)^2 \right] \left[1+\left(\frac{f}{8~ \rm{mHz}} \right)^4  \right],
    \label{eq:acc}
\end{equation}
\begin{equation}
    S_{\text{oms}}(f)=\left( s_{\text{oms}} \frac{2\pi f}{c} \right)^2 \left[1+\left(\frac{2~\rm{mHz}}{f} \right)^4 \right].
    \label{eq:oms}
\end{equation}
For simplicity in our simulation, we treat the functional forms of these PSD as identical across all six optical benches.
The noise magnitude parameters adopted for this study are set to $s_{\text{acc}}=3\times 10^{-15}~{\text{m}}/{\text{s}^2}/\sqrt{\text{Hz}}$ and $s_{\text{oms}}=8\times 10^{-12}~\text{m}/\sqrt{\text{Hz}}$.

\subsection{Data stream}
The total measured data, represented by the Doppler frequency shift $\eta_{\text{rs}}(t)$ along a single interferometer link, is the superposition of the SGWB signal and the dominant instrumental noise components \cite{du2025enhancingtaijisparameterestimation}
\begin{equation}
    \eta_{\text{rs}}(t) = y_{\text{rs}}(t) + n^{\text{acc}}_{\text{rs}}(t) + D_{\text{rs}}(t)n^{\text{acc}}_{\text{sr}}(t) + n^{\text{oms}}_{\text{rs}}(t).
\end{equation}
Here, $D_{\text{rs}}(t)$ is defined as the time-delay operator, which shifts the signal by $L_{\text{rs}}(t)$
\begin{equation}
    D_{\text{rs}}(t)h(t)=h\left(t-L_{\text{rs}}(t)\right).
\end{equation}

In space-based gravitational-wave detectors, fluctuations in the laser frequency noise constitute the dominant source of instrumental noise. Due to the unequal and time-varying arm-lengths arising from the orbital motion of the spacecraft, this noise cannot be canceled through simple interference, unlike in ground-based detectors. To overcome this challenge, the post-processing technique known as TDI has been developed to effectively suppress the laser frequency noise, as detailed in Refs.~\cite{Prince_2002,Tinto_2003,Tinto:2020fcc} and references therein. 
In this work, we adopt the first-generation Michelson configurations $X$, $Y$ and $Z$, under the condition that the laser frequency noise has been perfectly removed. The total measured signal in the 
$X$ channel can then be expressed as a linear combination of the single-link observables
\begin{equation}
    \begin{split}
        X(t)&=\left[\eta_{12}(t)+D_{12}\eta_{21}(t)-D_{13}D_{31}\eta_{12}(t)-D_{13}D_{31}D_{12}\eta_{21}(t)\right] \\
        &-\left[\eta_{13}(t)+D_{13}\eta_{31}(t)-D_{12}D_{21}\eta_{13}(t)-D_{12}D_{21}D_{13}\eta_{31}(t)\right].
    \end{split}
\end{equation}
The corresponding expressions for the $Y(t)$ and $Z(t)$ channels are obtained by cyclically permuting the numerical subscripts $1\to 2\to 3 \to 1$ in the above formula.

\begin{figure}[tbh]
    \centering
    \includegraphics[width=12cm, height=9cm]{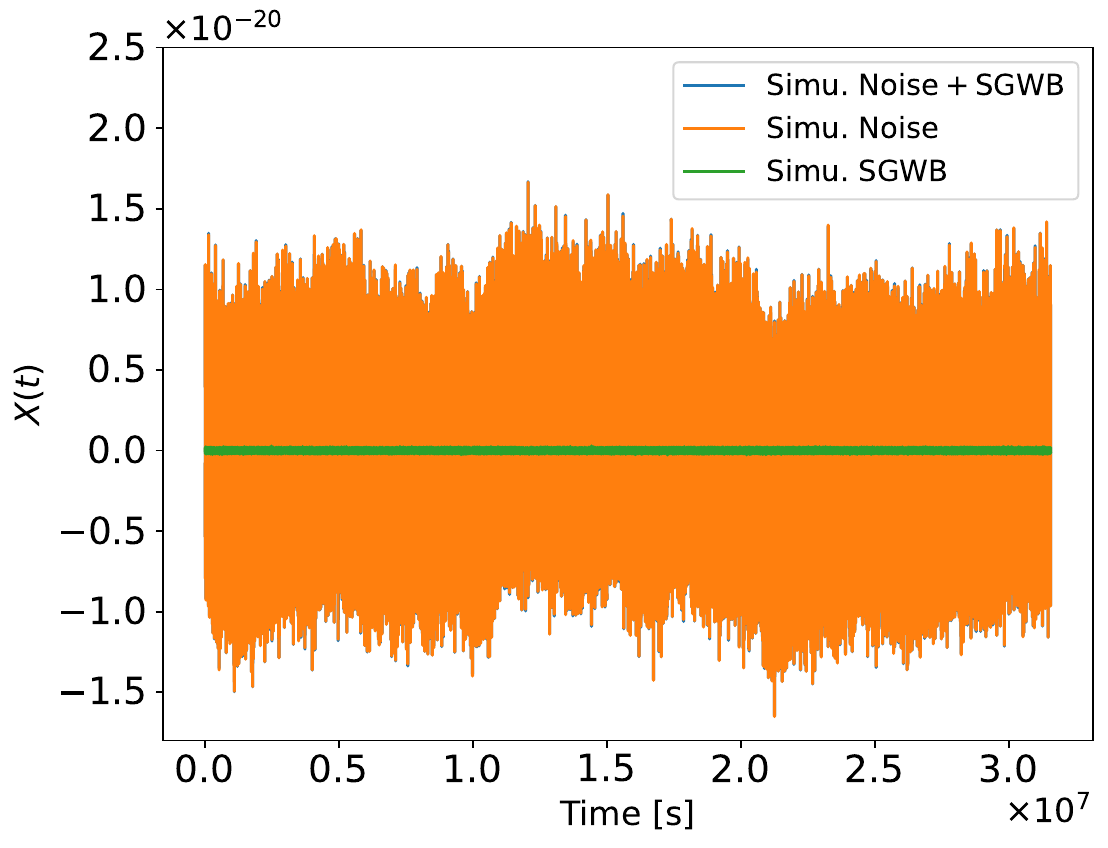}
    \caption{Simulated time series of $X$ channel. The data includes the SGWB (Eq.~\ref{eq:gwx}) and the combined instrumental noise (Eq.~\ref{eq:noisex}). The time series spans a duration of one year and is sampled at a frequency $f_s=0.1~\text{Hz}$. With the adopted SGWB parameters, $\Omega_0=3\times10^{-12}$ and $\gamma=-1$, the SNR at $X$ for the total observation period is approximately 79.}
    \label{fig:simu_time}
\end{figure}

We generate the time-domain data in $XYZ$ channels by utilizing $\textbf{Triangle-Simulator}$~\cite{du2025realisticdetectionpipelinestaiji}. In Fig.~\ref{fig:simu_time}, we present the simulated data of $X$ channel over a duration of one year, sampled at a frequency $f_s=0.1~\text{Hz}$. For comparison, we also plot the simulated SGWB corresponding to an amplitude of $\Omega_0=3\times10^{-12}$ and a spectral index $\gamma=-1$. As expected, the SGWB signal cannot be reliably disentangled from the instrumental noise components in the time domain, but only be inferred with PSD. The visual overall ratio of the strength between noise and SGWB is dominated by the large high-frequency components in oms noise. As we shall demonstrate in the following analysis, working in the frequency domain provides a much clearer view of the relative size of each component across the different frequency modes, which is essential for accurate parameter estimation.

Following the data generation, we calculate the PSD using the Welch method. The total data stream is divided into 146 segments, each corresponding to a duration of $T_n = 2.5$ days. To suppress spectral leakage, a Kaiser window with a shape parameter $\beta=28$ is applied.
Due to the sensitive band of the detector and the numerical instability arising from oscillations in the high-frequency region of the TDI response functions \cite{babak2021lisasensitivitysnrcalculations}, we truncate the analysis to the frequency range $[10^{-4},2\times10^{-2}]~\text{Hz}$.
The signal-to-noise ratio (SNR) for the injected SGWB at $X$ is calculated to be $\text{SNR}_{X}\simeq79$, defined as \cite{Caprini_2019}
\begin{equation}
    \text{SNR}_X=\sqrt{T_{\text{obs}}\int_{f_{\text{min}}}^{f_{\text{max}}}\left[\frac{S_{\text{gw},X}(f)}{S_{n,X}(f)}\right]^2\dd{f}},
\end{equation}
where $S_{\text{gw},X}$ and $S_{n,X}$ denote the PSD of the SGWB and instrumental noise at $X$, respectively. These PSD can be expressed as the product of the TDI response functions and the PSD of the three constituent sources
\begin{equation}
    S_{\text{gw},X}(f)=\tilde{R}_{\text{gw},X}(f)S_{\text{gw}}(f),
    \label{eq:gwx}
\end{equation}
\begin{equation}
    S_{n,X}(f)=\tilde{R}_{\text{acc},X}(f)S_{\text{acc}}(f)+\tilde{R}_{\text{oms},X}(f)S_{\text{oms}}(f).
    \label{eq:noisex}
\end{equation}

\begin{figure}[tbh]
    \centering
    \includegraphics[width=12cm, height=9cm]{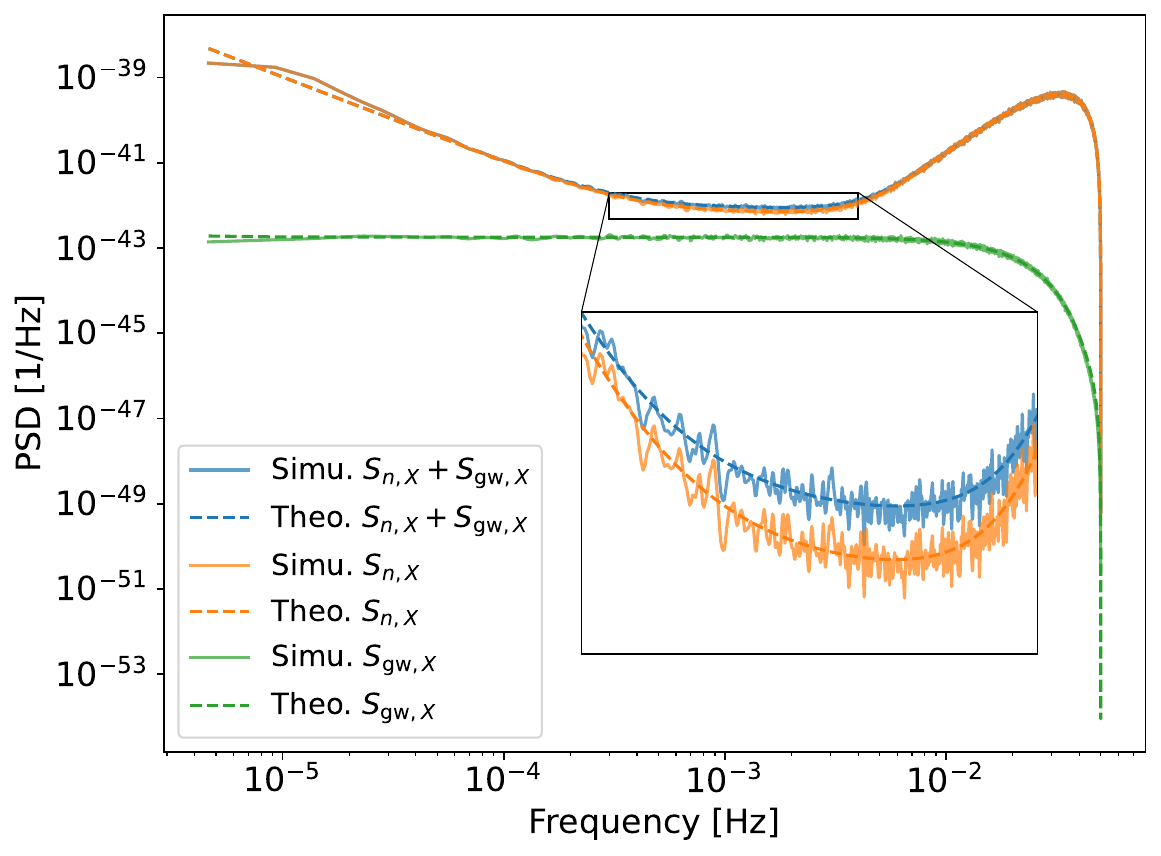}
    \caption{Comparison of the PSD of the simulated data and the theoretical model at $X$. The figure explicitly shows the contribution of the SGWB (Eq.~\ref{eq:gwx}) and the combined instrumental noise (Eq.~\ref{eq:noisex}) in comparison with the total observed PSD.}
    \label{fig:simu_psd}
\end{figure}

Fig.~\ref{fig:simu_psd} presents a comparison between the estimated PSD of the simulated time series and the corresponding theoretical model. The theoretical curves for the SGWB and instrumental noise components are calculated using the source PSD models detailed in Eq.~\ref{eq:singlepw}, Eq.~\ref{eq:acc}, and Eq.~\ref{eq:oms}, combined with the TDI response functions calculated under the idealized equal-arm model, which are listed in the subsequent section. As clearly demonstrated in the figure, the simulated PSD exhibits excellent agreement with the theoretical prediction across the entire frequency band. We note that, in our simulations, the total data stream is generated directly as the sum of the SGWB and instrumental noise components. The individual SGWB and noise components are plotted separately only to confirm the correct injection levels and validate the reliability of the simulation module.

\section{Parameter Inference} \label{sec:data_analysis}
In this section, we establish the Bayesian parameter inference framework used for our analysis. We detail the construction of the likelihood function and derive the necessary response functions in frequency domain for both the SGWB and instrumental noise, which form the basis of the analysis templates employed in the parameter estimation process.

\subsection{Likelihood Function}
Given the frequency-domain data $\vb{\tilde{d}}_{\eta}(f)=\mathcal{F}\left[\vb{d}_{\eta}(t)\right], \vb{d}_{\eta}(t)=\left\{X(t), Y(t), Z(t) \right\}$, 
we employ Bayes' theorem to perform maximum a posterior (MAP) estimation of the model parameters
\begin{equation}
    p\left(\vb{\theta}|\vb{\tilde{d}}_{\eta}\right)\propto \mathcal{L}\left(\vb{\tilde{d}}_{\eta}|\vb{\theta}\right)p(\vb{\theta}),
\end{equation}
where $p\left(\vb{\theta}|\vb{\tilde{d}}_{\eta}\right)$ is the posterior distribution, $\mathcal{L}\left(\vb{\tilde{d}}_{\eta}|\vb{\theta}\right)$ the likelihood function, and $p(\vb{\theta})$ the prior distribution of the parameters. 
The model parameter vector is defined as $\vb{\theta}=\left\{\log_{10}\Omega_0, \gamma, s_{\text{oms}}, s_{\text{acc}}\right\}$, with the corresponding prior distributions listed in Table~\ref{tab:prior}.
\begin{table}
\caption{Prior distributions for the parameters. The parameters $\log_{10}\Omega_0$ and $\gamma$ specify the amplitude and the spectral slope of the single power-law SGWB model Eq.~\ref{eq:singlepw}, while $s_{\text{oms}}$ and $s_{\text{acc}}$ denote the amplitudes of the noise spectra in Eq.~\ref{eq:acc} and Eq.~\ref{eq:oms}. The symbol $\mathcal{U}\left[\text{min}, \text{max}\right]$ represents a uniform distribution spanning the given range.}
\begin{ruledtabular}
\begin{tabular}{lccr}
$\quad$&Parameters & Prior&$\quad$ \\
\hline
$\quad$&$\log_{10}\Omega_0$ & $\mathcal{U}\left[-15, -5 \right]$&$\quad$ \\
$\quad$&$\gamma$ & $\mathcal{U}\left[-3, 3 \right]$&$\quad$ \\
$\quad$&$s_{\text{oms}}~\text{pm}/\sqrt{\text{Hz}}$ & $\mathcal{U}\left[5, 20\right]$&$\quad$ \\
$\quad$&$s_{\text{acc}}~{\text{fm}}/{\text{s}^2}/\sqrt{\text{Hz}}$ & $\mathcal{U}\left[0, 10\right]$&$\quad$ \\
\end{tabular}
\end{ruledtabular}
\label{tab:prior}
\end{table}

We use the Whittle likelihood for the distribution of the frequency spectrum at discrete sampling frequencies $f_k$ \cite{Karnesis_2020}
\begin{equation}
    \ln\mathcal{L}\left(\vb{\tilde{d}}_{\eta}|\vb{\theta}\right)=-\frac{N_c}{2}\sum_{f_k} \left[\vb{\tilde{d}}_{\eta}(f_k)  \vb{\Sigma}^{-1}(f_k;\vb{\theta}) \vb{\tilde{d}}_{\eta}^{\top}(f_k) + \ln\abs{\vb{\Sigma}(f_k;\vb{\theta})} \right].
\end{equation}
While some studies address potential biases arising from non-Gaussian data by employing alternative forms of the likelihood \cite{Flauger_2021}, we adopt the standard Gaussian likelihood function here. Such a correlated likelihood has been widely used for both ground-based detectors and space-based networks \cite{Zhu_2013,Fan_2018,Jiang_2023,Allen_1999,Romano:2016dpx,Cornish_2001,Cai_2024}.
In this expression, $N_{c}$ denotes the number of segments used in the PSD estimation, which is set to 146 in this work. The matrix $\vb{\Sigma}$ represents the covariance matrix of $\vb{\tilde{d}}_{\eta}$, with $\vb{\Sigma}^{-1}$ and $\abs{\vb{\Sigma}}$ denoting its inverse and determinant, respectively. 
Using the PSD definition in Eq.~\ref{eq:psd_sum}, the covariance matrix $\vb{\Sigma}$ takes the form
\begin{equation}
    \vb{\Sigma}=\frac{T_{\text{obs}}}{2} \begin{bmatrix}
S_{X} & S_{XY} & S_{XZ} \\
S_{YX} & S_{Y}& S_{YZ} \\
S_{ZX} & S_{ZY} & S_{Z} \\
\end{bmatrix}.
\end{equation}
The diagonal elements of $\vb{\Sigma}$ correspond to the total PSD in each TDI channel, while the off-diagonal elements represent the cross-spectral densities (CSD) between pairs of TDI channels. They are formally defined as the ensemble averages
\begin{equation}
    S_{X}(f_k)=\frac{2}{T_{\text{obs}}}\expval{\tilde{X}(f_k)\tilde{X}^*(f_k)}_{\text{ens}}=S_{\text{gw},X}(f_k)+S_{n,X}(f_k),
\end{equation}
\begin{equation}
    S_{XY}(f_k)=\frac{2}{T_{\text{obs}}}\expval{\tilde{X}(f_k)\tilde{Y}^*(f_k)}_{\text{ens}}=S_{\text{gw},XY}(f_k)+S_{n,XY}(f_k).
\end{equation}
The expressions for the other channels are defined analogously using their corresponding data. 
Finally, the sampling of the posterior distribution $p\left(\vb{\theta}|\vb{\tilde{d}}_{\eta}\right)$ is carried out via nested sampling, which is efficient for exploring and estimating the evidence of models in high-dimensional parameter spaces.

\subsection{Time-averaged template}
In the parameter estimation process, accurate frequency-domain template is required to characterize and match the theoretical predictions to the observed data. Given the known PSD models for both the SGWB and instrumental noise, the corresponding TDI response functions, $\tilde{R}_{\text{gw},X}(f)$, $\tilde{R}_{\text{acc},X}(f)$, and $\tilde{R}_{\text{oms},X}(f)$, must be rigorously derived. We obtain these essential response functions using three distinct methodologies, which are detailed in the remainder of this subsection.

The orbital motion of the spacecraft introduces time-dependent arm-lengths, making the response non-stationary and complicating the computation of a separable frequency-domain response.
Since the PSD is estimated using the Welch method by averaging spectra across 146 short segments, a natural approach is to compute the frequency-domain responses for each segment and then take the mean to construct the template. 
Within each short-time segment of duration, the spacecraft positions and arm-lengths vary slowly and can therefore be treated as quasi-static. 
For the $i$-th segment, corresponding to the time interval $\left[(i-1)T_n, iT_n\right]$, we select the segment center $t_i=\frac{2i-1}{2}T_n$ as the
reference time for the satellites' positions and the arm-lengths. 
In this setting, the final time-averaged frequency-domain response for the SGWB at $X$ is computed as
\begin{equation}
    \tilde{R}_{\text{gw},X}(f)=\frac{1}{N_c}\sum_{i=1}^{N_c}\tilde{R}_{\text{gw},X}(f;t_i).
    \label{eq:avrgresp}
\end{equation}
The corresponding response functions for the instrumental noise components and those in the other TDI channels can be obtained similarly.

\begin{figure}[t]
    \centering
    \includegraphics[width=11cm, height=10cm]{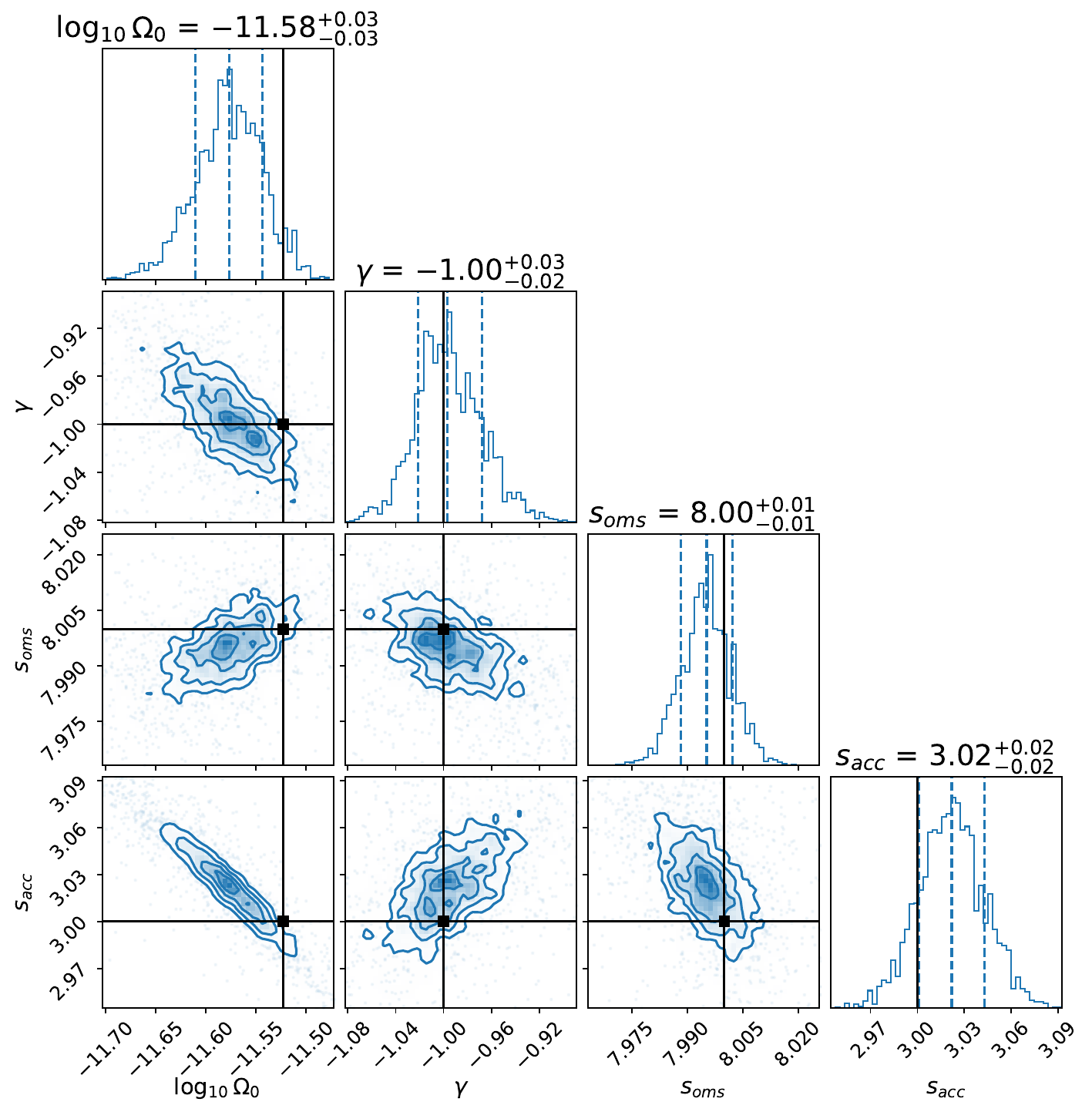}
    \caption{Marginalized posterior distributions for the model parameters $\vb{\theta}$ obtained using time-averaged frequency-domain template. The black solid line shows the injected parameter values. The vertical lines in the diagonal panels show the median and the $68\%$ CI. The contour lines in the off-diagonal panels represent the standard $1\sigma$, $2\sigma$, and $3\sigma$ iso-probability-density levels.}
    \label{fig:corner}
\end{figure}

\begin{figure}[t]
    \centering
    \includegraphics[width=10cm, height=8cm]{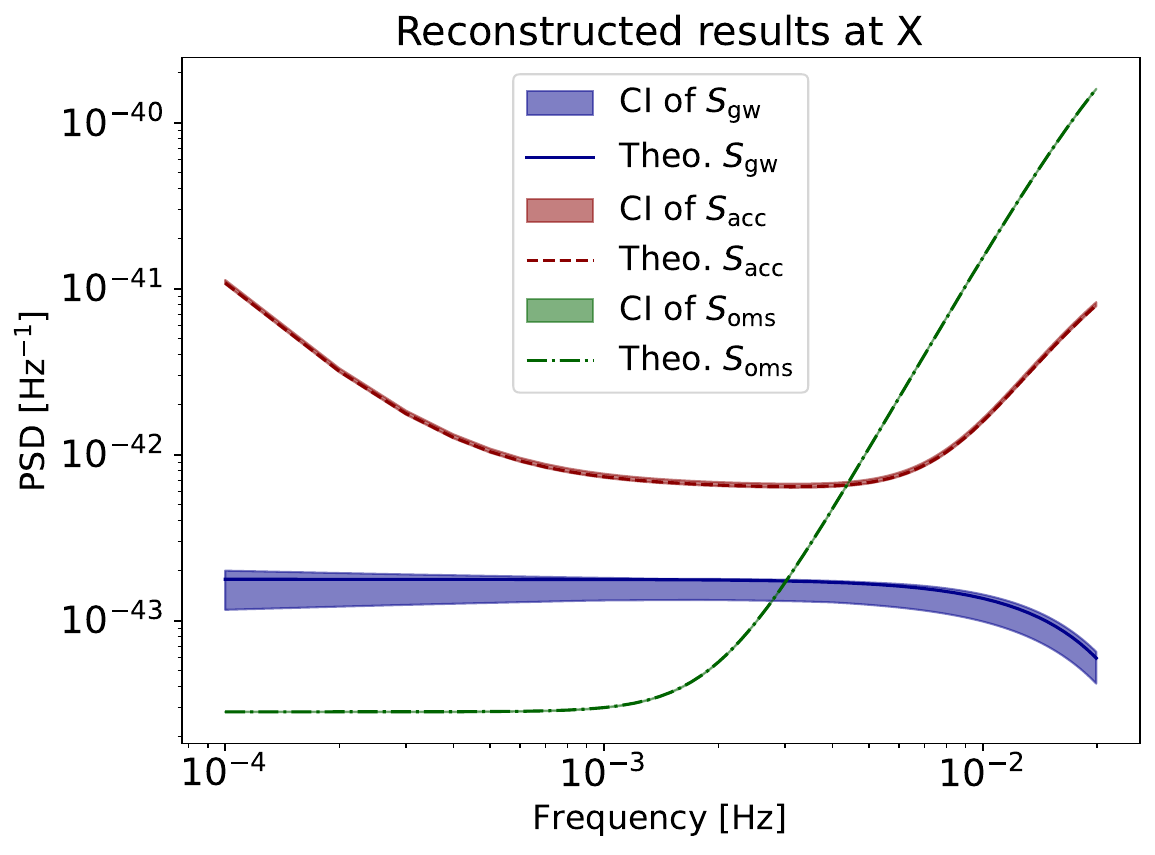}
    \caption{The reconstructed PSD for the SGWB and the instrumental noise at $X$. 
    The solid blue curve denotes the theoretical PSD of the SGWB, the dashed red curve represents the acceleration noise, and the dash-dotted green curve corresponds to the optical metrology system noise. 
    The lightly shaded regions in the respective colors indicate the 95$\%$ CI.}
    \label{fig:recover}
\end{figure}

To assess the effectiveness of the time-averaged template in characterizing both the signal and noise, we conduct parameter estimation on simulated data.
As plotted in Fig.~\ref{fig:corner}, the marginal posterior distributions show that most recovered parameters fall within their respective $68\%$ credible intervals (CI). However, we note a slight bias in the recovery of the SGWB amplitude, $\log_{10}\Omega_0$. This minor deviation from the injected value is likely attributable to the inherent stochastic nature of the data realization combined with the challenges of high-precision parameter estimation. To further validate the physical fidelity of our recovery, we plot the the reconstructed PSD curves in Fig.~\ref{fig:recover}. The correspondence between the reconstructed PSD and the theoretical model is evident, with the theoretical curve falling comfortably within the $95\%$ CI shown by the shaded regions. This close agreement demonstrates the high accuracy of the parameter estimation process.

To facilitate comparison and provide an analytical benchmark, we also derive the TDI template under the idealized equal-arm configuration. In this model, the arm-lengths are treated as constant and equal, $L_{\text{rs}}(t)=L=3\times10^9~\text{m}/c$.
While the SGWB responses are computed numerically using the dedicated simulation toolkit, the templates for PSD and CSD of the instrumental noise are derived from established analytical expressions \cite{Hartwig_2023}
\begin{equation}
    S_{n,X}=S_{n,Y}=S_{n,Z}=16\sin^2\left(2\pi fL \right) \left[\left(3+\cos\left(4\pi fL \right) \right)S_{\text{acc}}+S_{\text{oms}} \right],
    \label{eq:snx}
\end{equation} 
\begin{equation}
    S_{n,XY}=S_{n,YZ}=S_{n,ZX}=-4\sin\left(2\pi fL \right)\sin\left(4\pi fL \right) \left(4S_{\text{acc}}+S_{\text{oms}} \right).
    \label{eq:snxy}
\end{equation}

\begin{figure}[t]
    \centering
    \includegraphics[width=12cm, height=12cm]{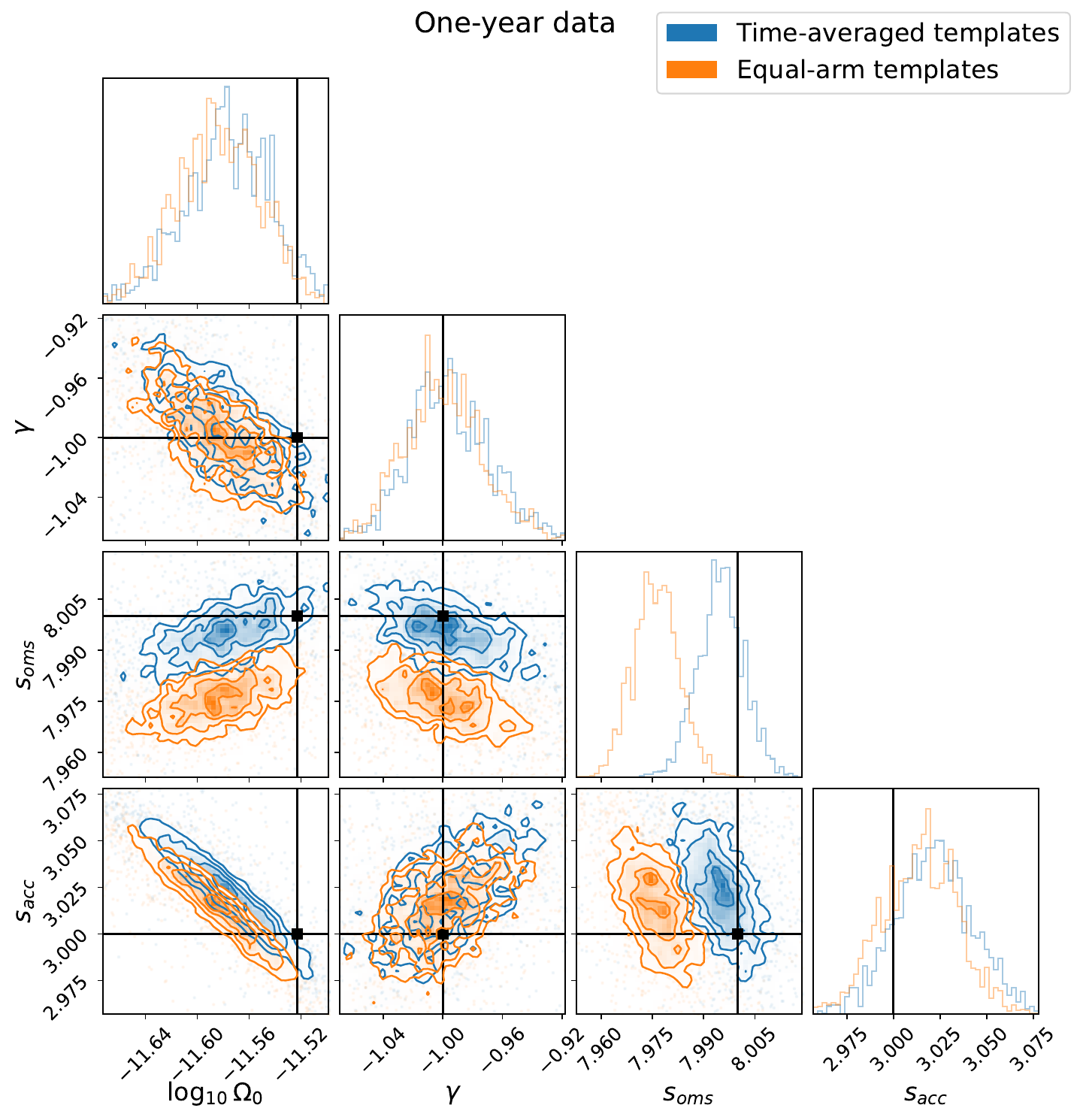}
    \caption{Marginalized posterior distributions obtained by fitting one year of simulated data using the time-averaged template and the equal-arm approximated template.}
    \label{fig:comparison}
\end{figure}

Fig.~\ref{fig:comparison} presents the parameter estimation results obtained from one-year data for the two distinct template cases.
From the panels, we observe that the bias between the injected values and the median of the posterior distribution for the oms amplitude $s_\text{oms}$ is notably larger when using the equal-arm template compared to the time-averaged template. In contrast, the bias remains almost negligible for the other parameters. This behavior is mainly attributed to the fact that the high-frequency oms noise dominates the PSD at $XYZ$. Since the high-frequency region also contains a larger number of sampled frequency points, the contribution of the oms noise to the overall likelihood becomes significantly enhanced. As a result, the inferred parameters are more sensitive to variations in the oms component of the template.

To further verify the advantage of the time-averaged template and evaluate their performance over varying observation duration, we additionally test the pipeline using half-year and three-month datasets while keeping the injected parameters fixed.
The number of segments used for PSD estimation must be carefully chosen, as spectral leakage can transfer power from the low-frequency range to higher frequencies. Given that the analyzed spectrum possesses significantly higher energy in the low-frequency band than in the mid-frequency band, it is consequently susceptible to these leakage effects. Excessive segmentation of the time series would shorten the effective duration of each segment and reduce the spectral resolution, thereby amplifying the leakage and degrading the estimation accuracy. To mitigate this issue, we divide the time series into $\lfloor N_c/m \rfloor$ segments, where the expression is rounded down to the nearest integer, and $m$ is the ratio between the one-year duration and the current observation time. 

\begin{figure}[htbp]
    \centering
    \includegraphics[width=11cm, height=10cm]{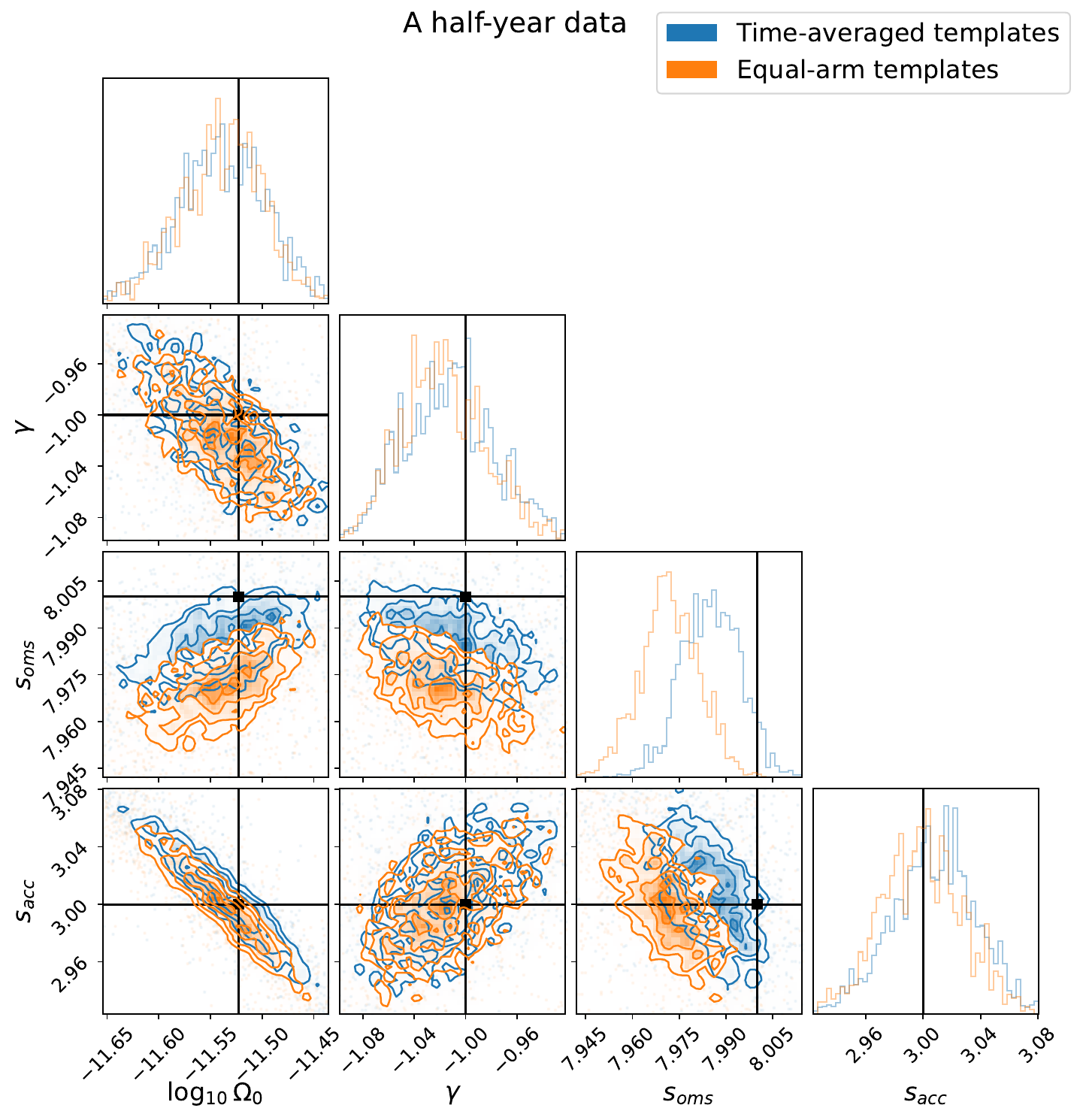}
    \includegraphics[width=11cm, height=10cm]{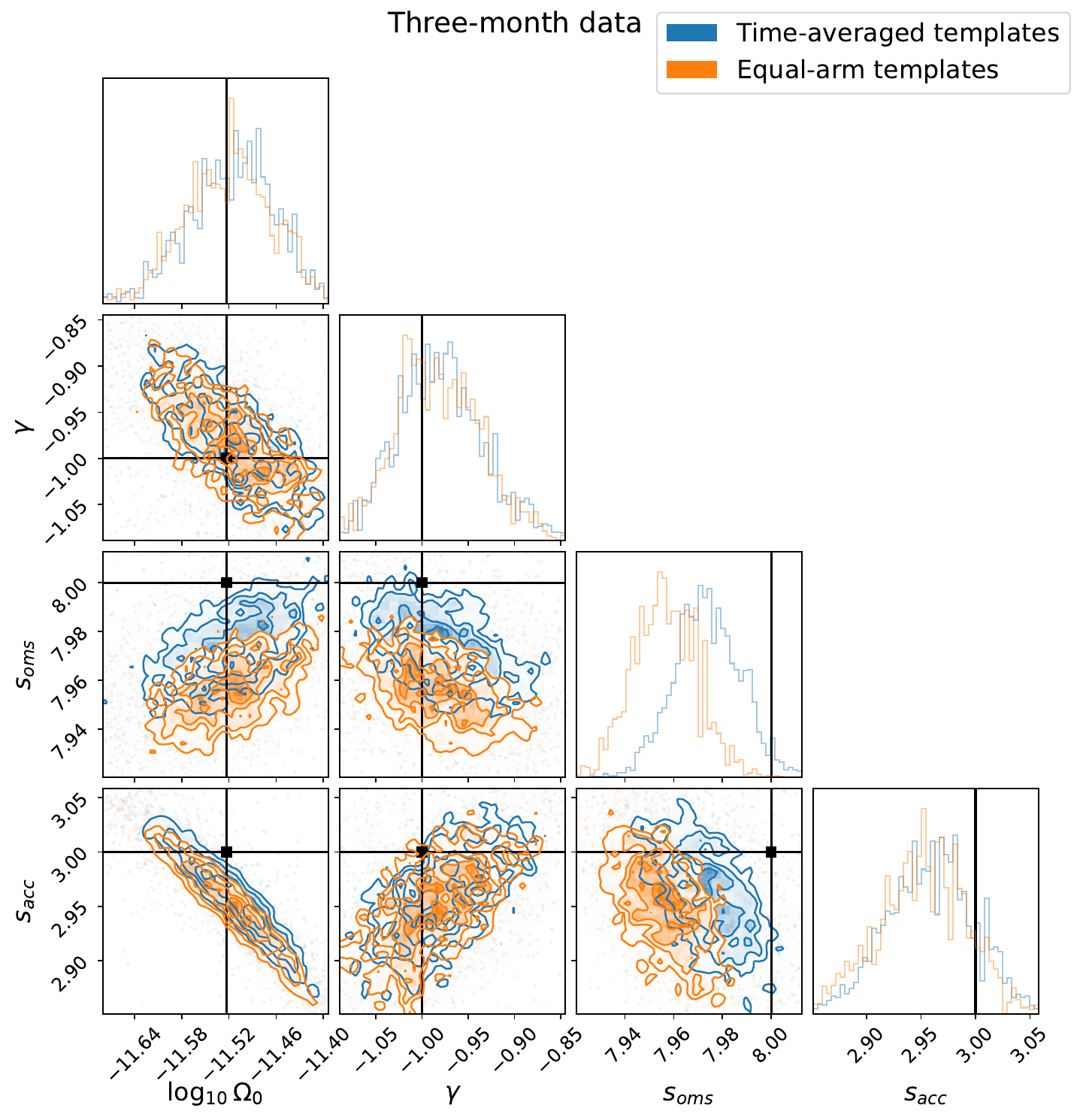}
    \caption{Marginalized posterior distributions obtained using the time-averaged and equal-arm template on shorter datasets: six months of simulated data (Top Panel) and three months of simulated data (Bottom Panel).}
    \label{fig:comparison_quatyear}
\end{figure}

In Fig.~\ref{fig:comparison_quatyear}, we present the comparison results for the half-year and three-month datasets, using $N_c=73$ and $N_c=36$ respectively. From these two analyses, we find that the injected value of the oms amplitude remains consistently closer to the median of the posterior distributions when using the time-averaged template, confirming their better performance over the equal-arm approximation. The combined results from all three tested datasets demonstrate that the time-averaged template effectively mitigate parameter discrepancies across varying integration times, thus confirming their necessity for accurate parameter estimation when modeling the detector's realistic orbit.

In addition to the comparative studies, we explore a scenario where the arm-length is treated as an unknown parameter within the inference process. We find that the time dependence in the noise responses arises primarily from the time-delay operator $D_{\text{rs}}(t)$. For the SGWB responses, an additional time dependence originates from the dynamic satellite positions and their relative orientations, as formally shown in Eq.~\ref{eq:gwfreshift}.
However, our tests indicate that the dominant effect of the dynamic orbit is due to variations in the arm-lengths, while the influence of satellite positions introduces only a relative error of $\mathcal{O}(10^{-5})$ in the SGWB responses. Therefore, the orbital modulation can be effectively captured by primarily accounting for changes in the arm-lengths.
During mission operation, the time-delay processes vary continuously but can be considered constant within a short-time segment, as implemented in the aforementioned time-averaged model. As an exploratory approach, we introduce an unknown effective arm-length $L_{\text{eff}}$ to capture the mean effect of the continuously varying arms. This approach replaces the fixed equal-arm length $L$ with $L_{\text{eff}} \times 10^9 ~\text{m}/c$, thereby treating the dynamic nature of the orbit as a single, free parameter within the parameter estimation framework.

Noted that, in reality, the six arms do not vary identically along the light-travel directions. This complexity, combined with the specific TDI data combination, implies that the theoretical noise response functions in Eq.~\ref{eq:snx} and Eq.~\ref{eq:snxy} should ideally incorporate different effective arm-lengths for the various delay terms present in the power spectra, similarly for the SGWB response function.
However, for this exploratory test, we treat $L_{\text{eff}}$ as a single parameter across all noise and signal components to examine the immediate effect of this increased model complexity on parameter estimation.
Since analytical expressions are not available for the SGWB responses under this scheme, we first simulate numerical equal-arm responses across a relevant range of arm-lengths. We then use interpolation on these results to efficiently obtain the SGWB responses for arbitrary values of $L_{\text{eff}}$ during the inference process. The prior distribution for $L_{\text{eff}}$ is taken to be uniform, $\mathcal{U}$, within the range $\left[2, 4\right]$.

\begin{figure}[htbp]
    \centering
    \includegraphics[width=11cm, height=10cm]{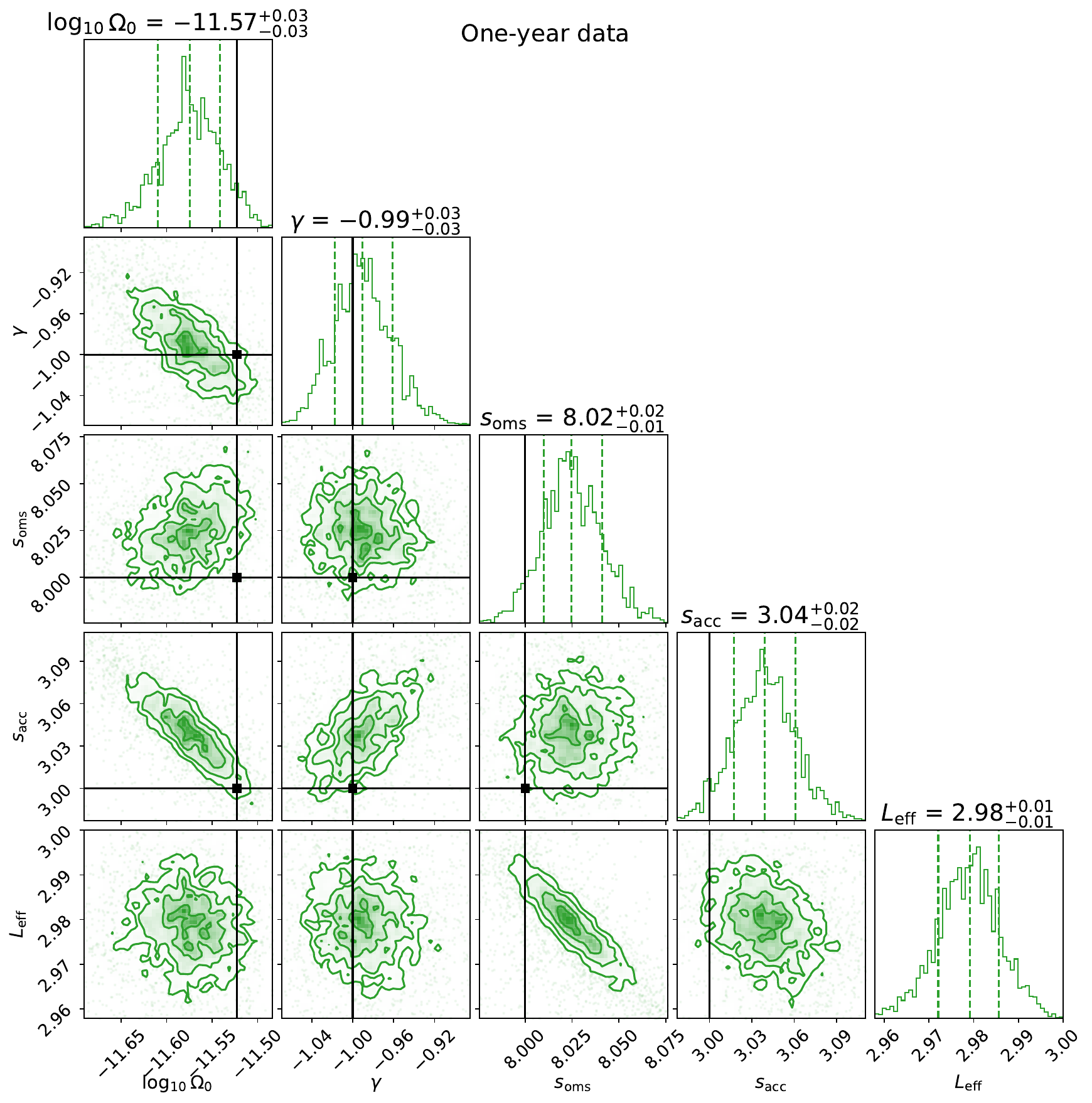}
    \caption{Marginalized posterior distributions obtained by treating the effective arm-length $L_{\text{eff}}$ as a free parameter, based on one year of simulated data. The true injected values for $L_{\text{eff}}$ are not displayed as this parameter is an unknown mean effect.}
    \label{fig:flexarm}
\end{figure}

\begin{figure}[htbp]
    \centering
    \includegraphics[width=11cm, height=10cm]{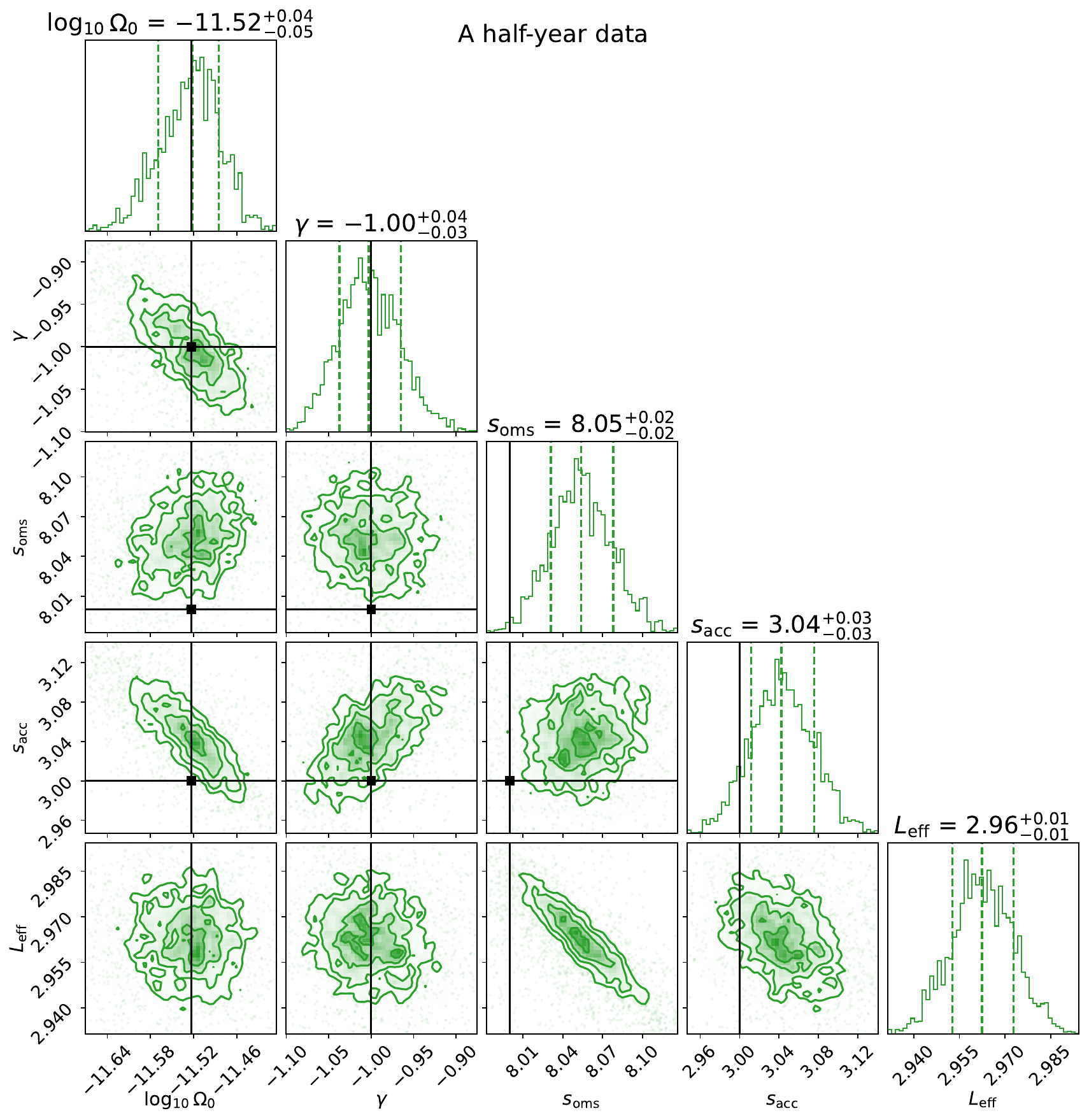}
    \includegraphics[width=11cm, height=10cm]{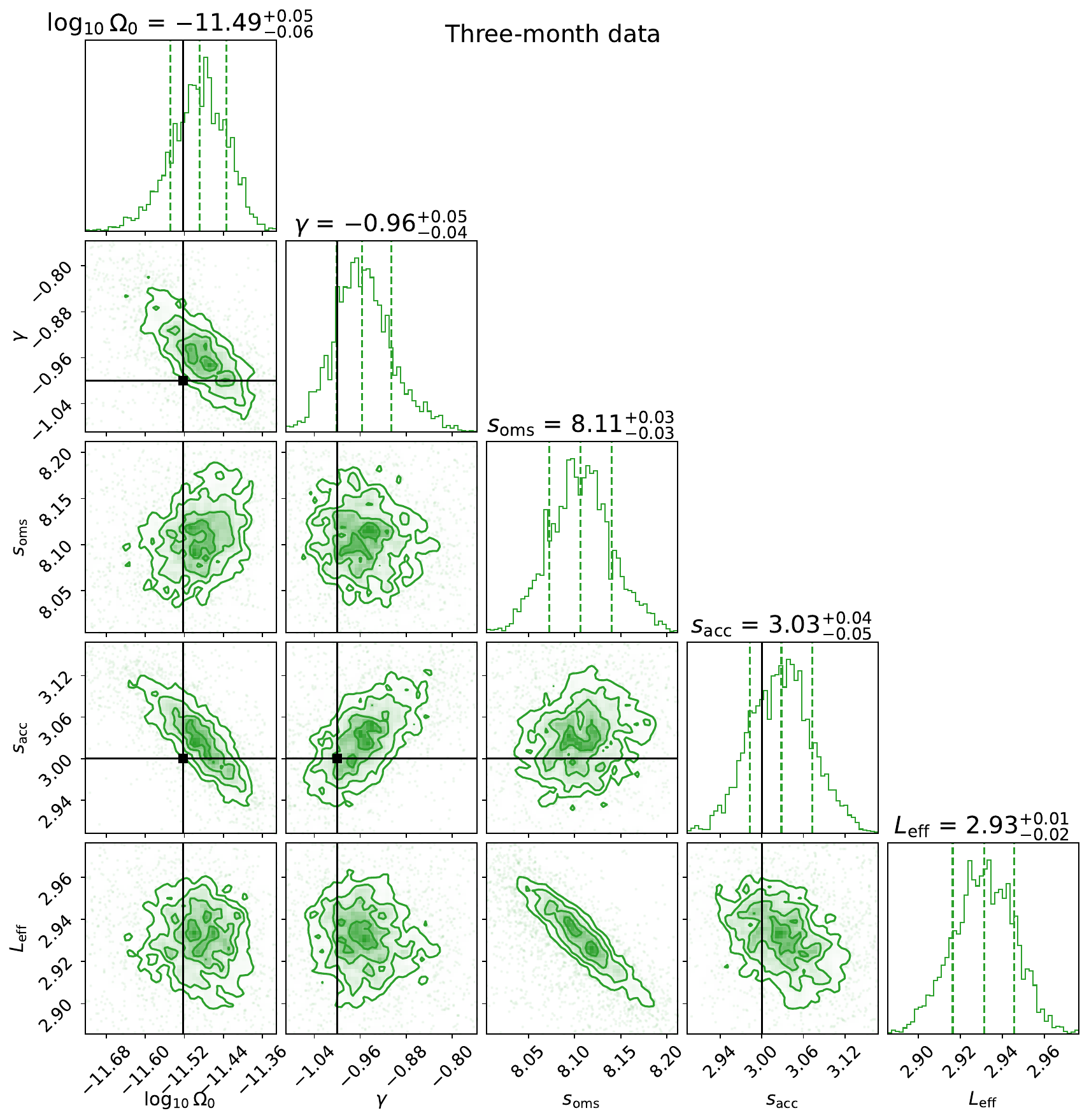}
    \caption{Marginalized posterior distributions obtained by treating the effective arm-length $L_{\text{eff}}$ as a free parameter, based on six months (Top Panel) and three months (Bottom Panel) of simulated data. The true injected values for $L_{\text{eff}}$ are not displayed as this parameter is an unknown mean effect.}
    \label{fig:flexarm_quatyear}
\end{figure}

We present the sampling results obtained using the unknown effective arm-length model in Fig.~\ref{fig:flexarm} and Fig.~\ref{fig:flexarm_quatyear}. Note that the true values of $L_{\text{eff}}$ are not displayed in the posterior distributions because $L_{\text{eff}}$ is an effective parameter introduced to capture the mean orbital effect and is therefore intrinsically unknown in the physical sense, rather than because the estimates deviate from a true, fixed value. Compared with the results from the fixed equal-arm template and the time-averaged template, the current results indicate that introducing the effective arm-length $L_{\text{eff}}$ leads to larger biases, particularly among the instrumental noise components. As evidenced in the figures, the injected true values for $s_{\text{oms}}$ and $s_{\text{acc}}$ are further away from the median of their respective posterior distributions. This suggests that the single parameter approximation, is not sufficient to accurately model the complex dynamic effects of the orbit within the TDI response functions.

\section{Bayes factor} \label{sec:results}

To evaluate how well different waveform models explain the observed data, we adopt the Bayes factor as the model selection criterion. Unlike parameter estimation where the posterior median may appear closer to the injected values for a certain template, the Bayes factor evaluates the full evidence of a model. It therefore provides an overall measure of how well a template captures the physical behavior encoded in the data. The Bayes factor between models $M_i$ and $M_j$ is defined as
\begin{equation}
    \ln K_{ij}\left(\vb{\tilde{d}}_{\eta}\right)=\ln Z\left(\vb{\tilde{d}}_{\eta}|M_{i}\right) - \ln Z\left(\vb{\tilde{d}}_{\eta}|M_{j}\right).
\end{equation}
$Z\left(\vb{\tilde{d}}_{\eta}|M_{i}\right)$ is the Bayesian evidence for model $M_i$, which is calculated with the marginal likelihood
\begin{equation}
    Z\left(\vb{\tilde{d}}_{\eta}|M_{i}\right) = \int \mathcal{L}\left(\vb{\tilde{d}}_{\eta}|\vb{\theta},M_i\right) p(\vb{\theta}|M_i) \dd{\vb{\theta}}.
\end{equation}
Here, $i \in \{1,2,3\}$ with ``1'' for the time-averaged template, ``2'' for the equal-arm template, and ``3'' for the template including $L_{\text{eff}}$. 

Table~\ref{tab:bayesfactor} presents the logarithmic Bayes factors for the three candidate models across datasets of different durations. The results indicate that, compared with the equal-arm template, the time-averaged template are generally preferred for data simulated with moving orbits. For the comparison between the time-averaged template and the $L_{\text{eff}}$ model, the three-month dataset shows an opposite preference. This discrepancy is likely due to the larger statistical fluctuations in the shorter dataset. For longer-duration datasets, which allow for more accurate PSD estimation, the Bayes factor increasingly favors the time-averaged template.
Crucially, the values of $\ln K_{32}$ suggest that incorporating $L_{\text{eff}}$ can better describe the data compared with the equal-arm template.

\begin{table}
\caption{Logarithmic Bayes factors comparing the three parameter estimation models: the time-averaged template ($M_1$), the equal-arm template ($M_2$), and the model including $L_{\text{eff}}$ ($M_3$), presented across datasets of different observation durations.}
\begin{ruledtabular}
\begin{tabular}{lccr}
$\quad$& $\ln K_{12}$ & $\ln K_{13}$ & $\ln K_{32}$ \\
\hline
One-year data & 5.74 & 5.89 & -0.14 \\
A half-year data & 6.32 & 2.95 & 3.37 \\
Three-months data & 2.91 & -3.72 & 6.63 \\
\end{tabular}
\end{ruledtabular}
\label{tab:bayesfactor}
\end{table}

\section{Conclusion and Discussion} \label{sec:conclusion}
In this study, we investigate the impact of different frequency-domain template construction methods on the accuracy of characterizing SGWB and instrumental noise, specifically under realistic orbital conditions with time-varying arm-lengths.
We compare three distinct approaches: the idealized equal-arm template, a model treating arm-length as a free parameter, and time-averaged template derived from segmented data.
Our results demonstrate that, for realistic orbits, the time-averaged template effectively reduces parameter estimation errors compared to the conventional equal-arm approximation, highlighting the inadequacy of the equal-arm simplification in high-precision analysis.
Conversely, increasing model complexity by introducing the arm-lengths as an additional free parameter leads to degraded estimation accuracy and larger uncertainties.
Nevertheless, the Bayes factor analysis suggests that accounting for arm-length variability remains important under time-dependent orbital configurations. 
Therefore, we conclude that for the analysis of long-duration data from missions operating in realistic orbits, segmenting the time series to construct time-averaged frequency-domain template provides a crucial and effective strategy for improving parameter estimation accuracy.

There remain several limitations to be addressed in future work. 
First, the present analysis assumes fixed-shape PSD for both the SGWB and instrumental noise, whereas in reality, these spectra may exhibit more complex or time-varying features. It is therefore necessary to extend the current framework to accommodate flexible spectral models or unequal noise levels across optical benches. 
Second, we represent the varying arm-lengths using a single effective parameter $L_{\text{eff}}$. As discussed in Section~\ref{sec:data_analysis}, the six light-travel directions experience different variations in arm-length. In future studies, each direction could be assigned its own effective arm-length to further test the validity of this method. 
Finally, this work considers only the two dominant and unavoidable noise sources. Additional instrumental noises, such as clock noise and tilt-to-length coupling, should be included in future analysis to more accurately reflect realistic experimental conditions.

\begin{acknowledgments}
This work is partly supported by the National Key Research and Development Program of China (Grant No.~2021YFC2201901), 
and the Fundamental Research Funds for the Central Universities. We acknowledge the use of \textbf{Triangle-Simulator}~\cite{du2025enhancingtaijisparameterestimation} and \textbf{dynesty}~\cite{sergey_koposov_2025_17268284}.
\end{acknowledgments}

\bibliographystyle{apsrev}
\bibliography{SGWB}

\end{document}